# A Newton method without evaluation of nonlinear function values

W. Chen

**Present mail address** (as a JSPS Postdoctoral Research Fellow): Apt.4, West 1$^{st}$ floor, Himawari-so, 316-2, Wakasato-kitaichi, Nagano-city, Nagano-ken, 380-0926, JAPAN

Permanent affiliation and mail address: Dr. Wen CHEN, P. O. Box 2-19-201, Jiangshu University of Science & Technology, Zhenjiang City, Jiangsu Province 212013, P. R. China

Present e-mail: chenw@homer.shinshu-u.ac.jp

Permanent email: chenwwhy@hotmail.com

**Abstract:**

The present author recently proposed and proved a relationship theorem between nonlinear polynomial equations and the corresponding Jacobian matrix. By using this theorem, this paper derives a Newton iterative formula without requiring the evaluation of nonlinear function values in the solution of nonlinear polynomial-only problems.

**Key words**. Newton method, Jacobian matrix, nonlinear polynomial-only problems.

## 1. Introduction:

Most of nonlinear algebraic systems of equations arise from the numerical discretization of differential or integral governing equations of physical and engineering problems. The finite element and finite difference methods are the standard techniques now in use for discretization. The size of the resultant analogue equations of these methods is usually very large. The Newton method and its variants are of central importance now to compute these nonlinear algebraic equations [1]. The total solution procedure often becomes very costly due to the large dimension of algebraic systems. The time-consuming effort lies in the iterative evaluation of function values, Jacobian matrix and its inverse. As was pointed out by Liu et al. [2], the repeated numerical integration of the stiffness matrices and force vectors, in the terminology of numerical analysis, the Jacobian matrix and nonlinear function values, significantly affects the efficiency and accuracy of the finite element nonlinear computation. An inordinate amount of computing time and storage requirements even prohibits such calculations.

Considerable research effort has been devoted to the development of some efficient nonlinear algorithms to reduce the cost in the evaluation of the Jacobian matrix and its inverse [1-3]. However, to the author's acknowledge, there is not effective approach available now in the literature to avoid or reduce the computing effort in the evaluation of the nonlinear function values. Although the computational burden in this aspect does not take a crucial part of nonlinear solutions, the relative cost is absolutely nontrivial in solution of a large nonlinear system especially for the finite element scheme. Very recently, the present author discovered and proved a relationship theorem between nonlinear polynomial equations and the corresponding Jacobian matrix [4, 5]. The objective of this note is to apply this theorem to derive a simple Newton iterative formula without requiring the evaluation of nonlinear function values.

## 2. Results

Recently, the present author [4, 5] proved the following theorem:

**Theorem 1.** If $N^{(m)}(U)$ and $J^{(m)}(U)$ are defined as nonlinear numerical analogue of the m order nonlinear differential operator and its corresponding Jacobian matrix, respectively, then $N^{(m)}(U) = \frac{1}{m} J^{(m)}(U)U$ is always satisfied irrespective if which numerical technique is employed to discretize.

Without loss of generality, consider a quadratic nonlinear algebraic equations

$$f(u) = Lu + N^{(2)}(u) + b = 0, \qquad (1)$$

where L is the linear coefficient matrix; $N^{(2)}(u)$ denotes the quadratic nonlinear terms and b represents the constant vector. u is the desired vector. The Newton iterative formula of equation (1) can be expressed as

$$u^{k+1} = u^k - J^{-1}(u^k) f(u^k), \qquad (2)$$

where u's with superscript k+1 and k mean respectively the iteration solutions at the (k+1)-th and k-th steps, and

$$J(u^k) = \frac{\partial f(u^k)}{\partial u} = L + J^{(2)}(u^k) \tag{3}$$

is the Jacobian matrix of nonlinear function f(u). $J^{(2)}(u)$ in the above equation is the Jacobian matrix of the quadratic nonlinear function vector $N^{(2)}(u)$. According to theorem 1, we have

$$\begin{aligned} f(u^k) &= Lu^k + \frac{1}{2}J^{(2)}(u^k)u^k + b \\ &= \frac{1}{2}\left[Lu^k + J^{(2)}(u^k)u^k\right] + \frac{1}{2}\left[Lu^k + 2b\right] \\ &= \frac{1}{2}J(u^k)u^k + \frac{1}{2}\left[Lu^k + 2b\right] \end{aligned} \tag{4}$$

Substituting equation (4) into equation (2) yields

$$u^{k+1} = \frac{1}{2}u^k - \frac{1}{2}J^{-1}(u^k)(Lu^k + 2b). \tag{5}$$

It is noted that the iterative formula (5) does not require the evaluation of the quadratic nonlinear function vector yet keeps the converging rate and accuracy of the Newton method.

For cubic nonlinear algebraic equation

$$Lu + N^{(3)}(u) = b, \tag{6}$$

where $N^{(3)}(u)$ represents the cubic nonlinear vector, we can derive the following formula

$$u^{k+1} = \frac{2}{3}u^k - \frac{1}{3}J^{-1}(u^k)(2Lu^k + 3b) \tag{7}$$

by using the same procedure for formula (5). Also, it is stressed that the above formula (7) is in fact equivalent to the standard Newton iterative formula (2) but need not calculate the cubic nonlinear function vector.

For a mixed quadratic and cubic nonlinear equation

$$Lu + N^{(2)}(u) + N^{(3)}(u) = b, \qquad (8)$$

we can derive the following formula

$$u^{k+1} = \frac{2}{3}u^k - \frac{1}{3}J^{-1}(u^k)\left(2Lu^k + N^{(2)}(u^k) + 3b\right). \qquad (9)$$

The evaluation of the cubic nonlinear $N^{(3)}(u)$ is avoided, while the calculation of quadratic nonlinear function $N^{(2)}(u)$ is required.

It is recognized that the numerical integration of force vectors (nonlinear function vector) is one of factors which influence the efficiency of some popular numerical techniques such finite element methods. The presented iterative formulas (5), (7) and (9) of the Newton method eliminate or alleviate the need of the evaluation of the force vectors. We have successfully tested the given formulas in the solution of the static Navier-Stokes equations of the driven cavity problem. The present iterative formulas may be especially useful to improve efficiency of various modified Newton methods, in which the repeated evaluation of the Jacobian and its inverse has been reduced or avoided. The cost in the calculation of function value occupies a larger part of the total computational effort [6].

It should also be pointed out that a very large class of real-world nonlinear problems can be modeled or numerically discretized polynomial-only algebraic system of equations.

The given iteration formulas (5), (7) and (9) are in general applicable for all these problems. Therefore, this work is of practical significance in broad physical and engineering areas.